\documentclass[preprint,aps,preprintnumbers,amsmath,amssymb,superscriptaddress]{revtex4}

\usepackage{graphicx}
\usepackage{dcolumn}
\usepackage{bm}

\begin{document}


\title{Photoluminescence of a microcavity quantum dot system in the quantum strong-coupling regime}

\author{Natsuko Ishida}
\address{Advanced Science Institute, RIKEN, Wako-shi, Saitama 351-0198, Japan}

\author{Tim Byrnes}
\affiliation{National Institute of Informatics, Chiyoda-ku, Tokyo 101-8430, Japan}

\author{Franco Nori}
\address{Advanced Science Institute, RIKEN, Wako-shi, Saitama 351-0198, Japan}
\address{Physics Department, University of Michigan, Ann Arbor, MI 48104-4313, USA}

\author{Yoshihisa Yamamoto}
\affiliation{National Institute of Informatics, Chiyoda-ku, Tokyo 101-8430, Japan}
\affiliation{E. L. Ginzton Laboratory, Stanford University, Stanford, CA 94305, USA}

\date{\today}

\begin{abstract}
The Jaynes-Cummings model, describing the interaction between a single two-level system and a photonic mode, 
has been used to describe a large variety of systems, ranging from cavity quantum electrodynamics, trapped ions, to superconducting qubits coupled to resonators.  Recently there has been renewed interest in studying the quantum strong-coupling (QSC) regime, where states with photon number greater than one are excited.  
This regime has been recently achieved in semiconductor nanostructures, where a quantum dot is trapped in a planar microcavity. 
Here we study the quantum strong-coupling regime by calculating its photoluminescence (PL) properties under a pulsed
excitation.  
We discuss the changes in the PL as the QSC regime is reached, which transitions between a peak around the cavity resonance to a doublet.  We particularly examine the variations of the PL in the time domain, under regimes of short and long pulse times relative to the microcavity decay time. 
\end{abstract}

\maketitle

Strong coupling is a phenomenon that has gained fundamental importance in the field of quantum technologies, where combinations of different systems are coupled quantum mechanically. Typically of 
such engineered systems, a matter field consisting of two levels is coupled to a photonic or phononic bosonic mode. 
For low densities, defined as being less than one excitation per two-level system,
the new excitations of the system are superpositions of the matter and bosonic fields. 
The ability to create a superposition between matter and bosonic fields is important in terms of quantum information processing applications
due to their ability to couple isolated quantum systems together.  
For example, a common bosonic mode may be used as a quantum communication channel between qubits, providing effective two-qubit interactions.
The Jaynes-Cummings Hamiltonian, which describes any two-level system interacting with a single bosonic mode, is given by 
\begin{eqnarray}
\mathcal{H}_{{\rm sys}}
=
\mathcal{H}_0
+
\mathcal{H}_{{\rm int}},
\label{eq:hamiltonian_polariton}
\end{eqnarray}
with
\begin{eqnarray}
\mathcal{H}_0
&=&
\hbar \omega \sigma^{+} \sigma^{-} + \left( \hbar \omega - \Delta \right) a^{\dagger} a, \\
\mathcal{H}_{{\rm int}}
&=&
g \left( \sigma^{-} a^{\dagger} + \sigma^{+} a \right),
\end{eqnarray}
where $\sigma^{\pm}$ are the raising and lowering operators for the two-level system, and 
$a^{\dagger}$ is the creation operator for the bosonic mode. The 
coupling between the photon and the two level system is given by $g$, the transition frequency of the two-level system is $ \omega $, and $ \Delta $ is the detuning between the bosonic mode and the qubit \cite{Scully1997, Walls1994,Laussy2009}.

At high densities, where there is more than one excitation per two-level system, the model predicts the quantum strong-coupling (QSC) regime. 
The quantum strong-coupling regime has been studied using superconducting qubits \cite{Fink2008,Ashhab2010,You2011}. 
This may be seen by directly diagonalizing the Hamiltonian, which may be easily performed since it preserves the total excitation number $ n_{{\rm tot}} = a^\dagger a  + | X \rangle \langle X | $. Considering the case of a resonant ($ \Delta = 0 $) qubit and cavity system, the eigenstates may be found to be
\begin{align}
\label{dressedstates}
|\pm,n \rangle = \frac{1}{\sqrt{2}} \left( |X \rangle | n-1 \rangle \pm |G \rangle | n \rangle  \right)
\end{align}
where $ | n \rangle = \frac{1}{\sqrt{n!}} (a^\dagger)^n | 0 \rangle $. 
The ground and excited states of the two-level system are denoted $ | G \rangle $ and $ | X \rangle $, respectively. The energies of the states are 
\begin{align}
E_n^\pm = n \hbar \omega \pm 2g \sqrt{n} 
\end{align}
thus for each excitation number $ n $ (with the exception of $ n = 0 $ which is unique) there are two solutions forming the Jaynes-Cummings ladder, with the energy splitting increasing with the excitation number $ n $.  
Recently an experiment studying the Jaynes-Cummings ladder in the quantum strong-coupling
regime was realized, where a single quantum dot was placed in a microcavity structure \cite{kasprzak10,Santori2009}. 
The highly excited states of the Jaynes-Cummings ladder were accessed by a combination of off-resonant
pumping which first excites the $ n_{{\rm tot}} = 1 $ manifold, then resonant pumping which results in climbing to higher 
$ n_{{\rm tot}} $ states. 

In this paper we perform a numerical study of the photoluminescence of the quantum dot-microcavity 
system.  While similar studies of the photoluminescence have been made for the Jaynes-Cummings model 
in a number of works \cite{Perea2004,Quesada2011,Gonzalez2010}, 
the pumping scheme that is used in these works typically consider a continuous-wave pumping.  
However, in order to achieve the high densities required for the QSC regime, 
experiments typically use a pulsed excitation due to the power constraints of the typical lasers used. Such 
pulsed excitations give a time-dependent, non-equilibrium element to the problem. The dynamics of the system 
is determined not only by the Hamiltonian (\ref{eq:hamiltonian_polariton}), but due to the open-dissipative
nature of the matter and bosonic fields, which have a finite decay time. 
In our calculations, we consider both regimes where the pulse
duration is shorter and longer than the lifetime of the particles in the system. 
Due to the pulsed excitation, the system is in general always in a non-equilibrium state. 
This allows for the possibility of directly studying the system during the decay process via observation of the PL characteristics. 

We also consider the effects of various pumping configurations, namely with respect to either the matter or bosonic fields \cite{Michler2000,Santori2002}. 
In general, it is also possible to observe the photoluminescence that is 
associated with either the matter or bosonic fields. 
In the case of experiments in Ref.~\cite{kasprzak10}, either the PL due to the photons leaking out of the cavity, or the PL due to recombination of the excitons can be distinguished. 
These can in principle give different spectral characteristics, as we find explicitly from our calculations. 
This gives four possible pumping/measurement schemes, the relevant one depends on the particular experimental configuration used. 
We calculate all four cases and find that there are particular spectral characteristics common to all configurations which give spectral evidence for the QSC regime being reached.

\section*{RESULTS}
\label{sec:dynamics}

\begin{figure}[t]
\begin{center}
\includegraphics[width=15cm,clip]{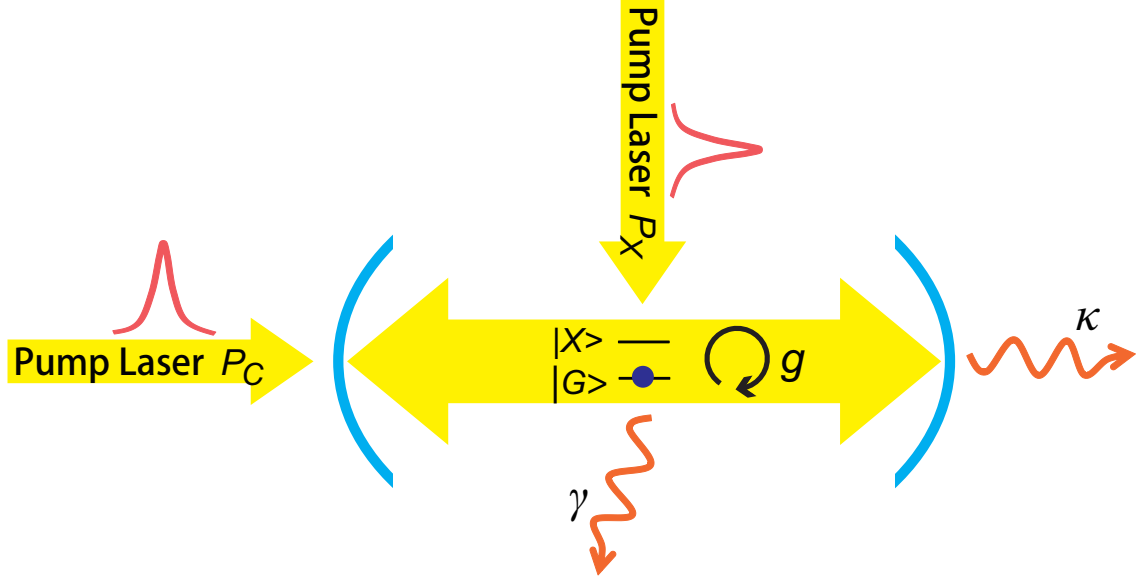}
\caption{The particles, oscillating between the ground state $|G\rangle$ and the excited state $|X\rangle$, are pumped in the cavity by a intense pump laser,  either $P_C$ or $P_X$. 
The excited particles in the cavity decay with either the exciton decay rate $\gamma$ or cavity photon decay rate $\kappa$. 
The coupling strength between the cavity photon and a two-level atom is given by $g$.}
\label{fig:Figure-1-Ishida}
\end{center}
\end{figure}

To model the open-dissipative dynamics of the coupled cavity-quantum dot system, we use a master equation
with the coherent part obeying (\ref{eq:hamiltonian_polariton}), Lindblad photon and exciton decay terms 
with rates  $\kappa$ and $ \gamma $ respectively, and Lindblad photon and exciton pump terms with 
rates $ P_C $ and $ P_X $ respectively (see Figure \ref{fig:Figure-1-Ishida}). 
The pump terms are assumed to obey a Gaussian profile in the time domain with time duration $ \tau_P $ (see Methods). 
The density matrix elements are numerically evolved in time, and the photoluminescence is calculated
using two-time correlation functions evaluated using the quantum regression theorem \cite{Perea2004,Quesada2011,Gonzalez2010}. 
A cutoff $ n_{{\rm max}} $ for the number of photons is made sufficiently large such that the trace of the density matrix is close to unity for all times.  For concreteness we have chosen parameters that are of the order of the experiment performed in Ref.~\cite{kasprzak10}. 
Specifically, the coupling constant $g$ is of the order of $10$ meV, while the excitation energy $\hbar \omega$ is of the order 
of eV; thus we fix $g=10$ and $\hbar \omega=1000$ where all the parameters are in units of meV. The detuning parameter is set to $\Delta=0$ throughout this study.

\subsection*{Density matrix dynamics}

\begin{figure}[t]
\includegraphics[width=15cm,clip]{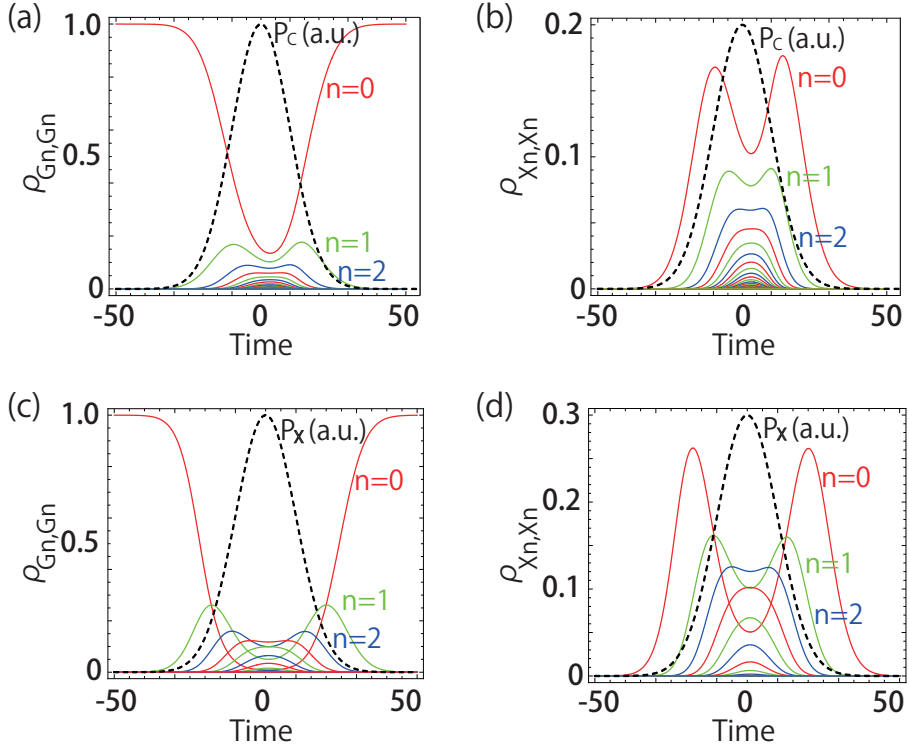}
\caption{Density matrix dynamics $\rho(t)$ for long pulse pumping with $\tau_P > 1/\kappa$. 
The time evolution of: 
(a) the occupation $\rho_{Gn,Gn}$ for the ground state $Gn$ under photon pumping $P_C$, 
(b) the occupation $\rho_{Xn,Xn}$ for the excited state $Xn$ under photon pumping $P_C$, 
(c) the occupation $\rho_{Gn,Gn}$ for the ground state $Gn$ under exciton pumping $P_X$, and 
(d) the occupation $\rho_{Xn,Xn}$ for the excited state $Xn$ under exciton pumping $P_X$. 
Curves are plotted in order of red ($n=0$), green ($n=1$), blue ($n=2$), red ($n=3$), green ($n=4$), blue ($n=5$), etc., with photon number $n$.
As references, the dashed curves indicate the pumping profile. 
Parameters used here are $g=10$, $\hbar \omega=1000$, $\tau_P=10$, $\gamma=0$, and $\kappa=1$. 
Negative times refer to prior to the pulse at $t=0$.}
\label{fig:Figure-2-Ishida}
\end{figure}

\begin{figure}[t]
\includegraphics[width=15cm]{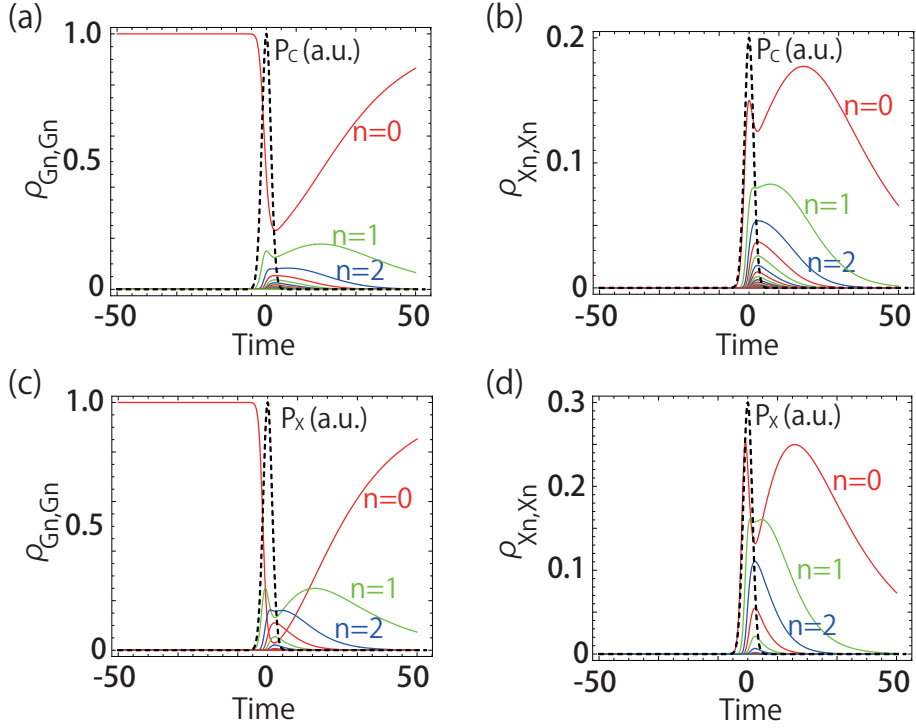}
\caption{Density matrix dynamics $\rho(t)$ for short pumping pulse with $ \tau_P < 1/\kappa $.
The time evolution of:
(a) the occupation $\rho_{Gn,Gn}$ for the ground state $Gn$ under photon pumping $P_C$, 
(b) the occupation $\rho_{Xn,Xn}$ for the excited state $Xn$ under photon pumping $P_C$, 
(c) the occupation $\rho_{Gn,Gn}$ for the ground state $Gn$ under exciton pumping $P_X$, and 
(d) the occupation $\rho_{Xn,Xn}$ for the excited state $Xn$ under exciton pumping $P_X$.
Curves are plotted in order of red ($n=0$), green ($n=1$), blue ($n=2$), red ($n=3$), green ($n=4$), blue ($n=5$), etc., with photon number $n$.
As references, the dashed curves indicate the pumping profile. 
Unlike the case with long pulse, the system evolves asymmetrically with time.
Parameters used here are $g=10$, $\hbar \omega=1000$, $\tau_P=1.5$, $\gamma=0$, and $\kappa=0.1$.
Negative times refer to prior to the pulse at $t=0$.}
\label{fig:Figure-3-Ishida}
\end{figure}

We consider two regimes in this study, where the pump timescale $\tau_P$ is either much longer or shorter than the 
polariton lifetime $ 1/\kappa $, when the system is pumped by either $P_C$ or $P_X$.  
Considering the case of long pumping $ \left( \tau_P > 1/\kappa \right)$ first, we show
the time evolution of the populations and the pump-pulse profile in Figure \ref{fig:Figure-2-Ishida}.
The pumping pulse is centered at $t_0=0$, and the system starts in the vacuum state $ | G\rangle | n=0 \rangle $.
We see that as the pump is increased, the probability that the state is in 
the vacuum state decreases, resulting in excitations being created.  
Cutoffs of $ n_{{\rm max}} =50 $ for $P_C$ and $ n_{{\rm max}} =30 $ for $P_X$ are sufficient for the density matrix to have $ \mbox{Tr} \rho = 1 $ for all times. 
At the strongest pumping intensity, there is a dip in some of the probabilities (e.g.  the $ | X\rangle | n = 1 \rangle $ state), signifying that excitations beyond the 
$n_{{\rm tot}} = 1 $ manifold are being excited.  This is by definition the QSC regime, and confirms that with 
suitably intense pumping such regimes may be reached.  

Similarly, Figure \ref{fig:Figure-3-Ishida} shows the density matrix elements for the case of a short pulse time $ \tau_P < 1/\kappa$.  
Again we see that many of the states above the $n_{{\rm tot}} = 1 $ manifold are being excited, confirming that either pumping scheme may be used to reach the QSC regime.  
We however see an asymmetric behavior to the density matrix elements,  unlike Figure \ref{fig:Figure-2-Ishida} which was approximately symmetric.  
This can be explained due to the relative timescales of the pumping, where either the system can respond fast enough 
to changes in the pumping (the former case), and the pump is faster than the timescale of the system (the latter case).  
In the case of the short pulse, the polaritons require a time $ \sim 1/\kappa $ before they can escape the system, 
thus the decay process itself is being observed for times $ t> t_0 $.


\subsection*{Photoluminescence spectra}
\label{sec:pl}

\begin{figure}[t]
\includegraphics[width=15cm]{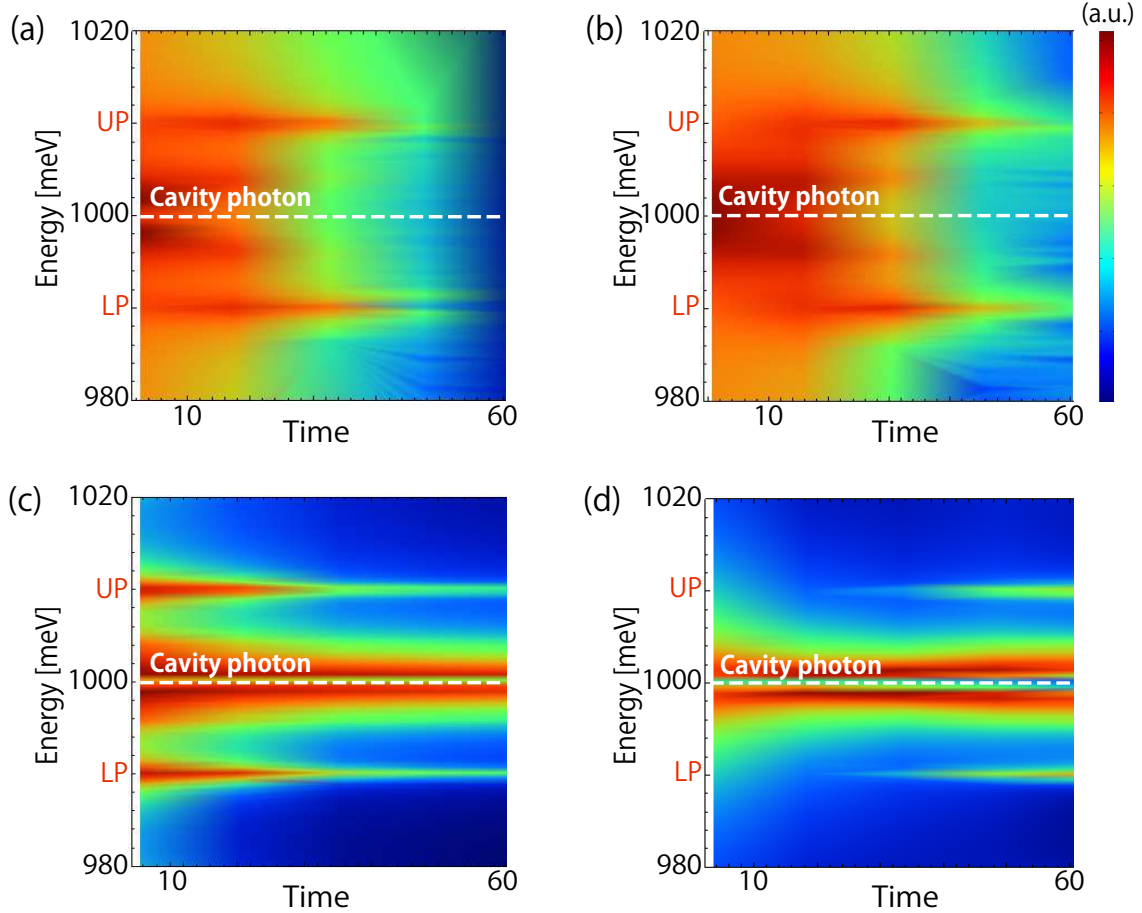}
\caption{Photoluminescence spectra for long pump pulses. 
The time evolution of:
(a) photonic PL by photon pumping $P_C$, 
(b) photonic PL by exciton pumping $P_X$, 
(c) excitonic PL by photon pumping $P_C$, and 
(d) excitonic PL by exciton pumping $P_X$ in the case of long pump pulse.
The peak PL approaches the cavity photon mode at high densities, whereas two peaks at UP and LP are brighter at low densities. 
Dashed line indicates the cavity photon energy. 
Parameters used are $g=10$, $\hbar \omega=1000$, $\tau_P=10$, and:
(a) $P_C^{{\rm max}}=0.8$, $\kappa=1$, $\gamma=0$, $n_{{\rm max}} =50$;  
(b) $P_X^{{\rm max}}=6.0$, $\kappa=1$, $\gamma=0$, $n_{{\rm max}} =30$; 
(c) $P_C^{{\rm max}}=0.3$, $\kappa=0$, $\gamma=1$, $n_{{\rm max}} =120$; and 
(d) $P_X^{{\rm max}}=3.5$, $\kappa=0$, $\gamma=1$, $n_{{\rm max}} =40$.
}
\label{fig:Figure-4-Ishida}
\end{figure}

\begin{figure}[t]
\includegraphics[width=15cm]{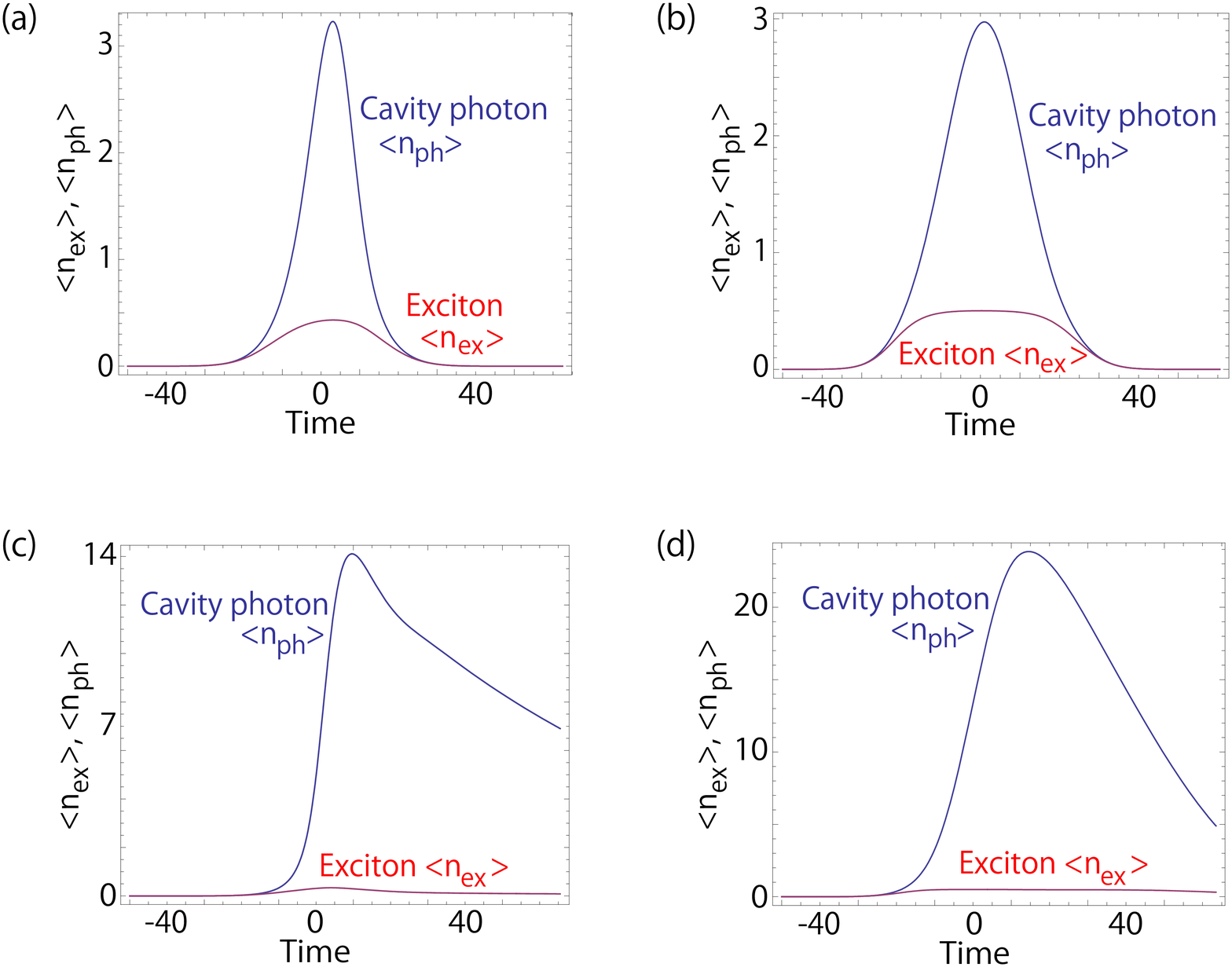}
\caption{The mean number of cavity photons $\langle n_{{\rm ph}} \rangle $ and excitons $ \langle n_{{\rm ex}} \rangle $ for long pump pulses. 
The cases shown are 
(a) photonic PL by photon pumping $P_C$, 
(b) photonic PL by exciton pumping $P_X$, 
(c) excitonic PL by photon pumping $P_C$, and 
(d) excitonic PL by exciton pumping $P_X$ in the case of long pump pulse. 
Parameters used are $g=10$, $\hbar \omega=1000$, $\tau_P=10$, and:
(a) $P_C^{{\rm max}}=0.8$, $\kappa=1$, $\gamma=0$, $n_{{\rm max}} =50$;  
(b) $P_X^{{\rm max}}=6.0$, $\kappa=1$, $\gamma=0$, $n_{{\rm max}} =30$; 
(c) $P_C^{{\rm max}}=0.3$, $\kappa=0$, $\gamma=1$, $n_{{\rm max}} =120$; and 
(d) $P_X^{{\rm max}}=3.5$, $\kappa=0$, $\gamma=1$, $n_{{\rm max}} =40$.
}
\label{fig:Figure-5-Ishida}
\end{figure}

We now show the photoluminescence dynamics of the quantum dot system.  We consider two cases of long pump pulses $\left( \tau_P > 1/\kappa, 1/\gamma \right)$ and short pump pulses $\left( \tau_P < 1/\kappa, 1/\gamma \right)$. 
For each case we evaluate the four cases of cavity photon pumping and exciton pumping with the measurement of the 
photonic or excitonic PL (Figure \ref{fig:Figure-4-Ishida} and \ref{fig:Figure-6-Ishida}). 
From the evolved matrix elements we also calculate the corresponding expectation values for the photon number $ n_{{\rm ph}} = a^\dagger a $ and the exciton number $ n_{{\rm ex}} = |X \rangle \langle X| $ (Figure \ref{fig:Figure-5-Ishida} and \ref{fig:Figure-7-Ishida}). 
We assume that the exciton decay rate $\gamma$ is negligible when measuring the photon mode, whereas the cavity photon decay rate $\kappa$ is negligible when measuring the exciton mode, such that $\gamma=0$ and $\kappa=0$ for photonic PL and excitonic PL, respectively. 

\subsubsection*{Photon PL and photon pumping}
First we examine the case where the system is photon pumped, and decays via photon emission $\kappa=1$. 
The average photon and exciton number is shown in Figure \ref{fig:Figure-5-Ishida}(a) for the long pump pulse case. 
We see that the photon number is much greater than the exciton number at the peak densities, which is a consequence of there being a maximum exciton number of 1, whereas there is no limit to the maximum photon number.  
For the quantum dot system this is a statement of Pauli's exclusion principle, as two excitons may not occupy the same dot (assuming the dot size is the same or smaller than the exciton Bohr radius).  
At high densities, there are therefore many more photons than excitons, again illustrating the QSC regime. 
For very high densities, such that $ \langle n_{{\rm ph}} \rangle \gg 1 $, the bosonic operators of the Hamiltonian (\ref{eq:hamiltonian_polariton}) may be written as a c-number, giving the effective Hamiltonian \cite{Byrnes2010}
\begin{align}
{\cal H}_{{\rm sys}} \approx \hbar \omega \langle n_{{\rm tot}} \rangle + g \sqrt{\langle n_{{\rm ph}} \rangle } \; \sigma^x
\end{align}
which has eigenstates $ \frac{1}{\sqrt{2}}\left( |X\rangle \pm |G\rangle \right) $.  Thus at high density the exciton population should be at most $ \langle n_{{\rm ex}} \rangle = 1/2 $. 
In Figure \ref{fig:Figure-5-Ishida}, we see that the exciton
number never exceeds this value, in agreement with this argument. 
 
In Figure \ref{fig:Figure-4-Ishida}(a) we see that at the highest densities the brightest peak comes from the cavity photon energy.  
As the system decays via leakage through the microcavity, the peak moves towards the lower polariton (LP) and upper polariton (UP) resonances, finally resulting in only the LP and UP resonances at the lowest density. A qualitative understanding of this spectrum can be described by using the dressed states $|\pm, n \rangle $ defined in equation (\ref{dressedstates}). The strong photon pumping initially creates a state which is 
\begin{eqnarray} 
\left( a^{\dagger} \right)^n |G \rangle |0\rangle
& \propto &
| G \rangle | n \rangle \nonumber \\
&=&
\frac{1}{\sqrt{2}}
\left(
| +, n \rangle -
| -, n \rangle
\right).
\label{photoninitialstate}
\end{eqnarray}
As the photon PL is observed due to a loss of a photon out of the system, this state transitions
to
\begin{align}
a |+, n \rangle &\approx \sqrt{n} \; |+, n-1 \rangle, \nonumber \\
a |-, n \rangle &\approx \sqrt{n} \; |-, n-1 \rangle,
\label{photonplphotonpump}
\end{align}
where we have assumed $ n \gg 1 $. The transition energy is then 
\begin{align}
\Delta E = E_n^{\pm} - E_{n-1}^{\pm} \approx \hbar \omega,
\end{align}
which corresponds to the cavity photon energy.  
In the low density limit, we have energy peaks at UP and LP as shown in Figure \ref{fig:Figure-4-Ishida}(a). 
Thus, we find that the energy at the cavity photon arises from the enhanced transitions of $|+,n\rangle \rightarrow |+,n-1\rangle$ 
and $|-,n\rangle \rightarrow |-,n-1\rangle$ due to the increased number of photons. 
The initial state has the same amplitude for the states $| +, n \rangle$ and $| -, n \rangle$, 
which is the reason for the symmetrical energy peaks to the cavity photon mode.

In Figure \ref{fig:Figure-7-Ishida} we plot the average number of photons and excitons for the short pump pulse case. 
We again see an asymmetric distribution of the particle numbers with respect to the maximum pump intensity, 
due to the slower decay time of the system with respect to the pump time. 
In all cases we again observe that the exciton number never exceeds 0.5 due to the same reasons as the long pump case. 

In Figure \ref{fig:Figure-6-Ishida} we show the PL spectrum for short pump pulses.  
We again see that at the highest densities the dominant peak is around the cavity photon energy.  As the system
decays away, there is small shift of the peak in the direction of the LP or UP energy.  At the lowest densities, only the 
LP or UP peak is visible. 
Apart from the decay dynamics, the PL signatures are fairly similar for both the long and short pump cases. 
In both cases we note that despite the high average densities achieved as seen in Figure 
\ref{fig:Figure-6-Ishida}, there is always a resonance peak visible at the LP and UP energies. 
According to the simplistic treatment in equation~(\ref{photonplphotonpump}) we would expect that these peaks would only be seen at the 
lowest densities. 
We see that in fact the dynamics is more complicated than the simplistic picture by 
again considering Figures \ref{fig:Figure-2-Ishida} and \ref{fig:Figure-3-Ishida}. If it were purely 
the case of (\ref{photonplphotonpump}) we would only see the density matrix elements corresponding to the highly
excited states and zero population of the low excitation states such as $ \rho_{G1,G1} $ and $ \rho_{X0,X0} $. 
In fact we see that for both the long and short pump-pulse cases, there is no reordering of the states, so that low 
excitation states are always more probable than the higher-excitation states. Thus there is always a strong contribution
of the transitions $ | \pm, 1 \rangle \rightarrow | 0 \rangle $ which are simply the LP and UP transitions. 
For this reason, for the typical pumping strengths that we consider, 
the PL spectrum consists of a mixture of both high and low excited states.  Only the parts of the PL spectrum 
in the vicinity of the cavity photon energy correspond to states in the QSC regime.


\subsubsection*{Photon PL and exciton pumping}

We now discuss the case of exciton pumping $P_X$ with photon decay, where we fix the decay rates to $\kappa=1$ and $\gamma=0$. 
The time evolution of the PL is shown in Figure \ref{fig:Figure-4-Ishida}(b) and the populations in Figure \ref{fig:Figure-5-Ishida}(b). 
In comparison to the photon pumping case, the system requires high pump powers to reach the high density due to the spin operator $\sigma^{+}$ suffering from phase-space filling, $(\sigma^{+})^n|0\rangle=0$ ($n \ge 2$). 
In order for the system to reach high densities for this pumping configuration, the pump must first create an exciton in the system, then this must
quickly become transfered to a photon via the coupling between the cavity and the quantum dot. 
If the pump can create another exciton in the system before the first photon decays, then high densities may be reached. 
The condition required for reaching the QSC regime is therefore $ P_X, g \gg \kappa $. 
Apart from this difference, we see a qualitatively similar spectrum to the photon PL/photon pumping case at high density, with the PL emerging most strongly at the cavity photon resonance. 
As the system decays, there is a shift of this peak towards the LP and UP resonances. 

For the short pump pulse case, we see an almost identical spectrum to the long pump-pulse case $\left[ \right.$Figures \ref{fig:Figure-6-Ishida}(b) and \ref{fig:Figure-7-Ishida}(b)$\left.\right]$. 
We can therefore conclude that despite the more inefficient pumping method of the exciton pumping, 
as long as the criterion $ P_X, g \gg \kappa $ is satisfied, the PL and population characteristics are independent of the particular 
pumping scheme that is used.


\subsubsection*{Exciton PL and photon pumping}

We now turn to the excitonic PL and photon pumping case and fix the decay rate $\kappa=0$ and $\gamma=1$.  
As seen in Figure \ref{fig:Figure-4-Ishida}(c), we observe a qualitatively different spectrum, 
with stronger side peaks at the LP and UP resonances, as well as the cavity photon resonance.  
We understand this by considering our simplified picture again. 
For strong photon pumping, we start once more in the state defined in equation (\ref{photoninitialstate}), 
$
(a^\dagger)^n |G \rangle | 0 \rangle \propto \frac{1}{\sqrt{2}}
\left(
| +, n \rangle -
| -, n \rangle
\right)
$. 
As the exciton PL corresponds to the loss of an exciton, we consider the transitions induced by 
operating $\sigma^{-}$ to the above states
\begin{align}
\sigma^{-}| \pm, n \rangle & = 
\frac{1}{2} \left( |+, n-1  \rangle - |-, n-1  \rangle \right)
\label{excitondecay}
\end{align}
Unlike the photon PL case in equation (\ref{photonplphotonpump}) where $ | \pm , n \rangle $ only transitions to one state, 
here there are two possible states to jump to. 
In this case all the transitions between $| \pm, n \rangle \rightarrow | \pm, n-1 \rangle$ have the same amplitude. 
Therefore, the most distinctive feature here is that we have side peaks at $\Delta E \approx \hbar \omega \pm 2g\sqrt{n}$, 
in addition to $\Delta E \approx \hbar \omega$ for high density. 
The kind of transitions shown above are familiar from resonance fluorescence, where in the high-excitation case the photoluminescence spectrum 
reduces to the Mollow's triplet spectrum  \cite{Mollow1969,Walls1994}. 
The difference here is that the densities are not quite so high, such that we may treat the excitation field as a classical field, 
and also the non-equilibrium nature of the pumping profile. 
Nonetheless, similar spectral characteristics may be seen for the high-excitation regime we consider here. 
For continuous and high-pumping rates we have confirmed that our numerics reproduces the Mollow's triplet spectrum.  

For short pump pulses, as shown in Figures \ref{fig:Figure-6-Ishida}(c) and \ref{fig:Figure-7-Ishida}(c), 
the numerical simulations are less stable with respect to the pumping dynamics, hence we were only able to achieve relatively
low densities (although still in the QSC regime) in comparison to the long pump-pulse cases. 
Nevertheless, we still see the characteristic Mollow's triplet spectrum evolving towards the LP and UP resonances. 
From this we conclude that
both the short and long pumping profiles should be effective in reaching the QSC regime, giving similar spectral characteristics.


\subsubsection*{Exciton PL and exciton pumping}

\begin{figure}[t]
\includegraphics[width=15cm]{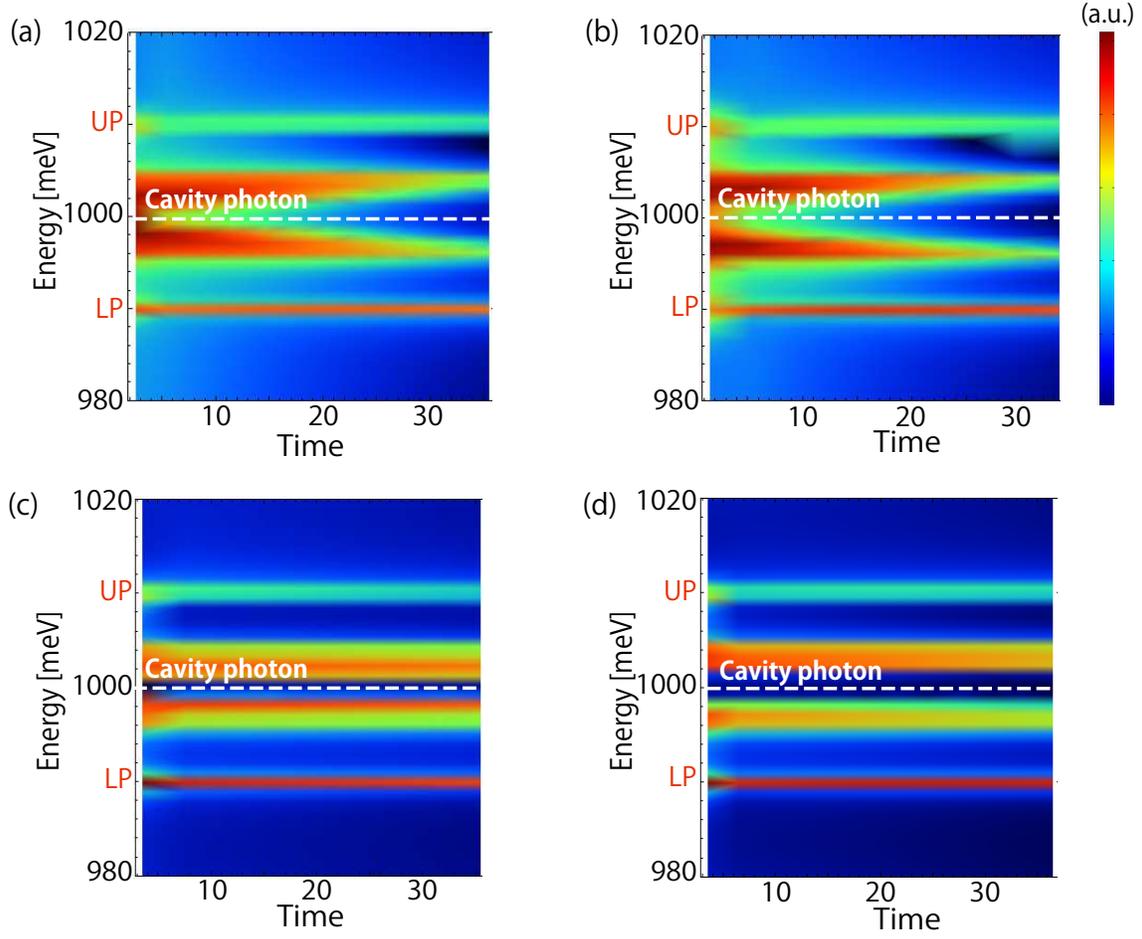}
\caption{Photoluminescence spectra for short pump pulses. 
The time evolution of:
(a) photonic PL by photon pumping $P_C$, 
(b) photonic PL by exciton pumping $P_X$, 
(c) excitonic PL by photon pumping $P_C$, and 
(d) excitonic PL by exciton pumping $P_X$. 
The peak PL approaches to the cavity photon mode at high densities, whereas two peaks at UP and LP are bright in addition to the peaks near the cavity photon resonance at low densities. 
Dashed line indicates the cavity photon energy. 
Parameters used are $g=10$, $\hbar \omega=1000$, $\tau_P=1.5$, and: 
(a) $P_C^{{\rm max}}=0.45$, $\kappa=0.1$, $\gamma=0$, $n_{{\rm max}} =40$; 
(b) $P_X^{{\rm max}}=1.3$, $\kappa=0.1$, $\gamma=0$, $n_{{\rm max}} =30$; 
(c) $P_C^{{\rm max}}=0.45$, $\kappa=0$, $\gamma=0.1$, $n_{{\rm max}} =40$; and 
(d) $P_X^{{\rm max}}=1.3$, $\kappa=0$, $\gamma=0.1$, $n_{{\rm max}} =40$.
}
\label{fig:Figure-6-Ishida}
\end{figure}

\begin{figure}[t]
\includegraphics[width=15cm]{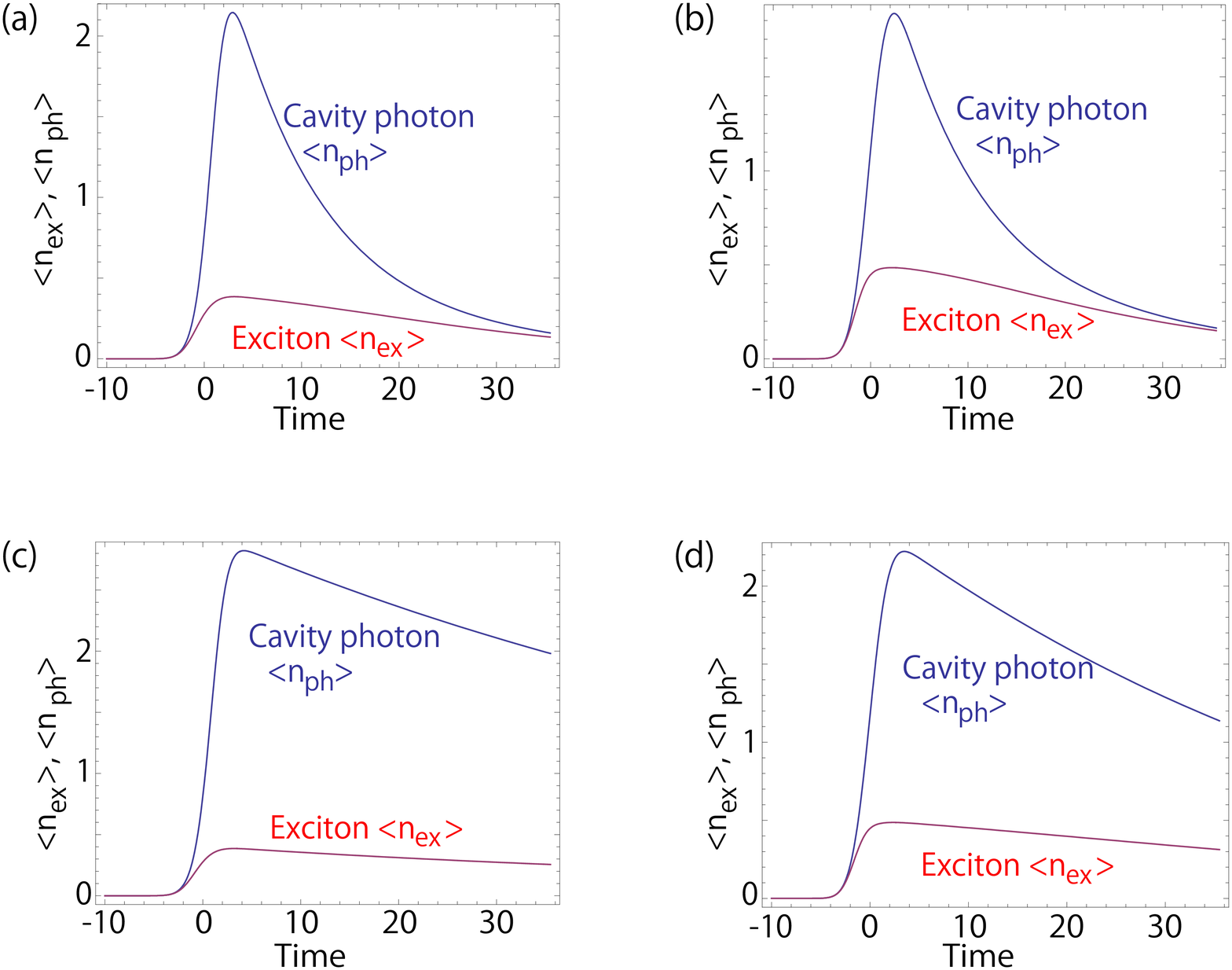}
\caption{The mean number of cavity photons $\langle n_{{\rm ph}} \rangle $ and excitons $ \langle n_{{\rm ex}} \rangle $ for short pump pulses. 
The cases shown are 
(a) photonic PL and photon pumping $P_C$, 
(b) photonic PL and exciton pumping $P_X$, 
(c) excitonic PL and photon pumping $P_C$, and 
(d) excitonic PL and exciton pumping $P_X$. 
Parameters used are $g=10$, $\hbar \omega=1000$, $\tau_P=1.5$, and:
(a) $P_C^{{\rm max}}=0.45$, $\kappa=0.1$, $\gamma=0$, $n_{{\rm max}} =40$; 
(b) $P_X^{{\rm max}}=1.3$, $\kappa=0.1$, $\gamma=0$, $n_{{\rm max}} =30$; 
(c) $P_C^{{\rm max}}=0.45$, $\kappa=0$, $\gamma=0.1$, $n_{{\rm max}} =40$; and 
(d) $P_X^{{\rm max}}=1.3$, $\kappa=0$, $\gamma=0.1$, $n_{{\rm max}} =40$.
}
\label{fig:Figure-7-Ishida}
\end{figure}

We finally consider the case where excitons are pumped and the exciton PL is measured with $\gamma=1$ and $ \kappa=0$. 
The time dependence of the excitonic PL is shown in Figures \ref{fig:Figure-4-Ishida}(d) and \ref{fig:Figure-6-Ishida}(d).
We see a qualitatively similar PL spectrum to the excitonic PL and cavity photon pumping case, with prominent side peaks on either side of the cavity photon resonance.
As was the case with photonic PL with excitonic pumping, the system requires 
high pump powers in order to reach the high-density regime, and requires $ P_X , g \gg \gamma $. 
In a similar way to the exciton PL and photon pumping case, all the four transitions in equation (\ref{excitondecay})
occur, giving rise to more prominent side peaks. 
We thus conclude that the particular form of the pumping from either short or long pulses, 
photonic or excitonic pumping gives qualitatively similar PL spectra for the excitonic PL cases.

\section*{DISCUSSION}
\label{sec:conc}

We have obtained the PL spectrum of a single quantum dot in a cavity pumped via injection of photons or excitons in a pulsed Gaussian profile. 
The particles decay either with the cavity photon decay rate $\kappa$ or with the exciton decay rate $\gamma$ depending on how we measure the PL, i.e. photonic PL or excitonic PL.
We first find that with any choice of pumping scheme the quantum strong-coupling regime can be reached for sufficiently strong pumping strengths. 
While this is intuitively clear for the photonic pumping case, this is also true for the exciton pumping, 
which suffers from the Pauli exclusion principle excluding it from the simultaneous injection of more than one exciton at a time. 
Despite this, the high density regime may be reached as long as the time required for conversion of excitons to photons 
via the coupling is faster than the decay rate $ P, g \gg \kappa, \gamma $. 
We also generally find that the PL spectrum that is measured is fairly insensitive to the particular pumping method used to excite the system. 
Our considerations of exciton and photon pumping, using both short and long profiles gives fairly consistent spectra throughout. 
Naturally, if the decay dynamics itself needs to be probed, then using a short pulse is really the only option. 
However, if the only goal is to reach the QSC regime, then either case can be used. 

The general trend of the photonic PL is that there is a peak at the cavity photon mode at high densities and as the system decays away it moves 
towards the UP and LP resonances in the low-density regime. 
For excitonic PL, the Mollow-triplet-like spectrum occurs as the density is increased. 
These properties agree with the theoretical predictions of \cite{Byrnes2010} at high density. 
Some differences to the PL spectrum is seen at intermediate densities (Figure 3(b) of Ref.~\cite{Byrnes2010}). 
In contrast to the many-exciton model, where the photoluminescence smoothly transitions from the LP to cavity photon energy, 
in the single quantum dot case the PL discontinuously shifts between the two limits. 
We have confirmed that by increasing the number of excitons in the system the transition gradually becomes smoother \cite{ishida13}. 
The discontinuous spectrum can therefore be seen as a signature of the quantum dot system, 
although in a typical experimental situation it is more likely that many quantum dots are excited simultaneously in order to have sufficient signal amplitude in the measurement.

\section*{METHODS}
\label{sec:plprop}

\subsection*{Time evolution of the density matrix}
Due to the open-dissipative nature of the quantum dot-microcavity system, 
the PL characteristics are modeled by a master equation following Refs. \cite{Perea2004,Quesada2011,Gonzalez2010}.  
We consider the reduced density matrix
\begin{eqnarray}
\frac{d}{dt}\rho
&=&
\frac{i}{\hbar}
\left[ \rho ,\mathcal{H}_{{\rm sys}} \right]
+
\frac{\kappa}{2}
\left ( 2 a \rho a^{\dagger} - a^{\dagger} a \rho - \rho a^{\dagger} a \right) 
\nonumber \\
&&+
\frac{\gamma}{2}
\left ( 2 \sigma^{-} \rho \sigma^{+} - \sigma^{+} \sigma^{-} \rho - \rho \sigma^{+} \sigma^{-} \right)
\nonumber \\
&&+
\frac{P_{C}(t)}{2}
\left( 2 a^{\dagger} \rho a - a a^{\dagger} \rho - \rho a a^{\dagger} \right)
\nonumber \\
&&+
\frac{P_{X}(t)}{2}
\left( 2 \sigma^{+} \rho \sigma^{-} - \sigma^{-} \sigma^{+} \rho - \rho \sigma^{-} \sigma^{+} \right),
\label{eq:master}
\end{eqnarray}
where we assume that the pump has a Gaussian time profile
\begin{eqnarray}
P_{C}(t),\; P_{X}(t)
=
P_0 \; {\rm exp}\left( -\frac{(t-t_0)^2}{2\tau_P^2}\right).
\end{eqnarray}

Using the basis $|i n\rangle$ where $i = G$ is the state with no exciton and $ i = X $ is the state with an exciton, 
we may define the matrix elements of the reduced density matrix
\begin{eqnarray}
\rho_{in, jm}
=
\langle
in | \rho | jm
\rangle,
\end{eqnarray}
where $ n, m $ is the photon number. The time evolution of the density matrix elements are
\begin{eqnarray}
\frac{d}{dt}\rho_{Gn, Gn}
&=&
\frac{i}{\hbar}g\sqrt{n} \left( \rho_{Gn, Xn-1}-\rho_{Xn-1, Gn} \right)
+
\gamma \rho_{Xn, Xn}
\nonumber \\
&&-\kappa \left[ n \rho_{Gn, Gn} - \left( n+1 \right) \rho_{Gn+1, Gn+1} \right] \nonumber \\
&&-P_X \rho_{Gn, Gn}
+P_C n \rho_{Gn-1, Gn-1} 
\nonumber \\
&&- P_C (n+1) \rho_{Gn, Gn}
\label{eq:population1}
,
\end{eqnarray}
\begin{eqnarray}
\frac{d}{dt}\rho_{Xn, Xn}
&=&
\frac{i}{\hbar}g\sqrt{n+1} \left( \rho_{Xn, Gn+1}-\rho_{Gn+1, Xn} \right)
-
\gamma \rho_{Xn, Xn}
\nonumber \\
&&-\kappa \left[ n \rho_{Xn, Xn} - \left( n+1 \right) \rho_{Xn+1, Xn+1} \right] \nonumber \\
&&+P_X \rho_{Gn, Gn}
+P_C n \rho_{Xn-1, Xn-1} 
\nonumber \\
&&- P_C (n+1) \rho_{Xn, Xn} 
\label{eq:population2}
,
\end{eqnarray}
\begin{eqnarray}
\frac{d}{dt}\rho_{Gn, Xn-1}
&=&
\frac{i}{\hbar} \left[ g \sqrt{n} \left( \rho_{Gn, Gn}-\rho_{Xn-1, Xn-1}  \right) \right. 
\nonumber \\
&& \left. \hspace{1em}
+\Delta \rho_{Gn, Xn-1} \right]\nonumber \\
&&- \left[ \left(
\gamma +\kappa(2n-1)+P_X
\right)/2 \right] \rho_{Gn, Xn-1}
\nonumber \\
&&+ \kappa \sqrt{n(n+1)} \rho_{Gn+1, Xn}\nonumber \\
&&+P_C \sqrt{n(n-1)} \rho_{Gn-1, Xn-2} 
\nonumber \\
&&- \frac{P_C}{2} (2n+1) \rho_{Gn, Xn-1} 
\label{eq:coherence1}
,
\end{eqnarray}
\begin{eqnarray}
\frac{d}{dt}\rho_{Xn-1,Gn}
&=&
-\frac{i}{\hbar} \left[ g \sqrt{n} \left( \rho_{Gn, Gn}-\rho_{Xn-1, Xn-1}  \right) \right. 
\nonumber \\
&& \left. \hspace{1em}
+\Delta \rho_{Xn-1,Gn} \right]\nonumber \\
&&- \left[ \left(
\gamma +\kappa(2n-1)+P_X
\right)/2 \right] \rho_{Xn-1, Gn}
\nonumber \\
&&+ \kappa \sqrt{n(n+1)} \rho_{Xn, Gn+1}\nonumber \\
&&+P_C \sqrt{n(n-1)} \rho_{Xn-2,Gn-1} 
\nonumber \\
&&- \frac{P_C}{2} (2n+1) \rho_{Xn-1,Gn} 
\label{eq:coherence2}
.
\end{eqnarray}
Imposing a maximum photon number $ n_{{\rm max}} $ we may evolve this set of equations which form 
a set of closed $ \sim 4 n_{{\rm max}} $ equations.  
In order to investigate the system dynamics, we truncate up to finite times instead of calculating up 
to the stationary limit.

\subsection*{Two-time correlation function}

In the quantum dot system, two types of photoluminescence mechanism can be considered: 
the particles are supplied from the pump laser and they decay due to leakage through 
the microcavity mirrors, or via exciton recombination \cite{Byrnes2010}. 
We call these processes the photonic and excitonic PL, respectively. 
The PL is given by \cite{Scully1997}
\begin{eqnarray}
S(r,\nu)
=
\frac{1}{\pi}
{\rm Re}
\int^{\infty}_{0}
\!\!
d \tau \; e^{i \nu \tau} \; G^{(1)}(t,\tau ).
\end{eqnarray}
where $ G^{(1)}(t,\tau ) $ is the two-time correlation function, which for the 
quantum dot system, is either
\begin{eqnarray}
G_{C}^{(1)}(t,\tau )
&=&
\langle a^{\dagger}(t+\tau ) a(t) \rangle, \\
G_{X}^{(1)}(t,\tau )
&=&
\langle \sigma^{+}(t+\tau ) \sigma^{-}(t) \rangle,
\end{eqnarray}
depending on which type of PL is being calculated. 
The Fourier transformation of $G_{C}^{(1)}(t,\tau )$ gives the photonic PL which is emitted by the leakage from the imperfect mirror, 
while that of $G_{X}^{(1)}(t,\tau )$ gives the excitonic PL which is emitted due to the recombination of excitons.

Following Refs.~\cite{Perea2004,Quesada2011,Gonzalez2010}, in the interaction picture we may conveniently define the following operators
\begin{eqnarray}
a^{\dagger}_{Gn}(\tau)
&=&
|G n+1 \rangle \langle Gn | e^{ i(\hbar \omega - \Delta) \tau/\hbar }, \nonumber \\
a^{\dagger}_{Xn}(\tau)
&=&
|X n+1 \rangle \langle Xn | e^{ i(\hbar \omega - \Delta) \tau/\hbar }, \nonumber \\
\sigma^{+}_{n}(\tau)
&=&
|Xn \rangle \langle Gn | e^{ i\hbar \omega \tau/\hbar }, \nonumber \\
\varsigma_{n}(\tau)
&=&
|G n+1 \rangle \langle Xn-1 | e^{ i(\hbar \omega - 2 \Delta) \tau/\hbar }.
\label{eq:operators}
\end{eqnarray}
In terms of the new operators defined in equation (\ref{eq:operators}), 
we can rewrite the two-time correlation function as
\begin{eqnarray}
G_{C}^{(1)}(t,\tau )
&=&
\sum_n
\sqrt{n+1}
\left (
\langle a_{Gn}^{\dagger}(t+\tau ) a(t) \rangle
+
\langle a_{Xn}^{\dagger}(t+\tau ) a(t) \rangle
\right), \nonumber \\
\\
G_{X}^{(1)}(t,\tau )
&=&
\sum_n
\langle \sigma_{n}^{+}(t+\tau ) \sigma^{-}(t) \rangle.
\label{eq:G1X}
\end{eqnarray}
In order to calculate the two-time correlation function, 
we need to obtain the expectation values of two operators at different times. 
The time evolution of the expectation values for each operators is then
\begin{eqnarray}
\frac{d}{d\tau} \langle a_{Gn}^{\dagger}(\tau ) \rangle
&=&
\langle a_{Gn}^{\dagger}(\tau ) \rangle \left[ \frac{i}{\hbar}( \hbar \omega-\Delta) -\frac{\kappa}{2}(2n+1)-P_X \right] \nonumber \\
&&+
\langle a_{Gn+1}^{\dagger}(\tau ) \rangle \kappa \sqrt{(n+1)(n+2)}
\nonumber \\
&&+
\langle \sigma_{n}^{+}(\tau ) \rangle \frac{i}{\hbar} g \sqrt{n+1}\nonumber \\
&&+
\langle a_{Xn}^{\dagger}(\tau ) \rangle \gamma
-
\langle \varsigma_{n}(\tau ) \rangle \frac{i}{\hbar} g \sqrt{n} \nonumber \\
&&+ 
\left[ \langle  a_{Gn-2}^{\dagger}(\tau ) \rangle \right] ^{\dagger} P_C \sqrt{n(n-1)}
\nonumber \\
&&-
\left[ \langle a_{Gn-1}^{\dagger}(\tau ) \rangle \right] ^{\dagger} \frac{P_C}{2}(2n+1), 
\label{eq:t_expect_operators1}
\end{eqnarray}
\begin{eqnarray}
\frac{d}{d\tau} \langle \sigma_{n}^{+}(\tau ) \rangle
&=&
\langle a_{Gn}^{\dagger}(\tau ) \rangle \frac{i}{\hbar} g \sqrt{n+1} 
\nonumber \\
&&+
\langle \sigma_{n}^{+} (\tau ) \rangle \left[ i \omega -\frac{(\gamma+P_X)}{2}-\kappa n \right]\nonumber \\
&&+
\langle \sigma_{n+1}^{+} (\tau ) \rangle \kappa (n+1) 
-
\langle a_{Xn-1}^{\dagger}(\tau ) \rangle \frac{i}{\hbar} g \sqrt{n} \nonumber \\
&&+
\langle \sigma_{n-1}^{+}(\tau ) \rangle P_C n
-
\langle \sigma_{n}^{+}(\tau ) \rangle P_C (n+1),
\label{eq:t_expect_operators2}
\end{eqnarray}
\begin{eqnarray}
\frac{d}{d\tau} \langle a_{Xn-1}^{\dagger}(\tau ) \rangle
&=&
\langle a_{Gn-1}^{\dagger}(\tau ) \rangle P_X
-
\langle \sigma_{n}^{+}(\tau ) \rangle \frac{i}{\hbar} g \sqrt{n} \nonumber \\
&&+
\langle a_{Xn-1}^{\dagger}(\tau ) \rangle \left[ \frac{i}{\hbar}( \hbar \omega-\Delta) -\gamma - \frac{\kappa}{2}(2n-1) \right] \nonumber \\
&&+
\langle a_{Xn}^{\dagger}(\tau ) \rangle  \kappa \sqrt{n(n+1)}
+
\langle \varsigma_{n}(\tau ) \rangle \frac{i}{\hbar} g \sqrt{n+1}\nonumber \\
&&+ 
\langle a_{Xn-2}^{\dagger}(\tau ) \rangle P_C \sqrt{n(n-1)}
\nonumber \\
&&-
\langle a_{Xn-1}^{\dagger}(\tau ) \rangle \frac{P_C}{2}(2n+1),
\label{eq:t_expect_operators3}
\end{eqnarray}
\begin{eqnarray}
\frac{d}{d\tau} \langle \varsigma_{n}(\tau ) \rangle
&=&
-\langle a_{Gn}^{\dagger}(\tau ) \rangle \frac{i}{\hbar} g \sqrt{n} 
+
\langle a_{Xn-1}^{\dagger}(\tau ) \rangle \frac{i}{\hbar} g \sqrt{n+1}\nonumber \\
&&+
\langle \varsigma_{n}(\tau ) \rangle \left[ \frac{i}{\hbar}( \hbar \omega-\Delta) -\frac{(\gamma+P_X)}{2}-\kappa n \right] 
\nonumber \\
&&+
\langle \varsigma_{n+1}(\tau ) \rangle \kappa \sqrt{n(n+2)}\nonumber \\
&&+
\langle \varsigma_{n-1}(\tau ) \rangle P_C \sqrt{(n-1)(n+1)}
\nonumber \\
&&-
\langle \varsigma_{n}(\tau ) \rangle P_C (n+1).
\label{eq:t_expect_operators4}
\end{eqnarray}
Applying the quantum regression formula to equations (\ref{eq:t_expect_operators1})-(\ref{eq:t_expect_operators4}),
we can compute and obtain the following sets of two-time correlation functions \cite{Carmichael2002}.
For the calculation of $G_{C}^{(1)}(t,\tau )$, we need to obtain
\begin{eqnarray}
\left\{
\langle a_{Gn}^{\dagger}(t+\tau) a(t) \rangle,
\langle \sigma_{n}^{+}(t+\tau ) a(t) \rangle,\right.\nonumber \\
\left.
\langle a_{Xn-1}^{\dagger}(t+\tau ) a(t) \rangle,
\langle \varsigma_{n}(t+\tau) a(t) \rangle
\right\}
\end{eqnarray}
and, for the calculation of $G_{X}^{(1)}(t,\tau )$, 
\begin{eqnarray}
\left\{
\langle a_{Gn}^{\dagger}(t+\tau) \sigma^{-}(t) \rangle,
\langle \sigma_{n}^{+}(t+\tau ) \sigma^{-}(t) \rangle,\right.\nonumber \\
\left.
\langle a_{Xn-1}^{\dagger}(t+\tau) \sigma^{-}(t) \rangle,
\langle \varsigma_{n}(t+\tau ) \sigma^{-}(t) \rangle
\right\}
\end{eqnarray}
is needed.
The above equations  satisfy the initial conditions
\begin{eqnarray}
\langle a_{Gn}^{\dagger}(t ) a(t) \rangle&=&\sqrt{n+1} \rho_{Gn+1,Gn+1}(t),\nonumber \\
\langle \sigma_{n}^{+}(t ) a(t) \rangle&=&\sqrt{n+1} \rho_{Gn+1,Xn}(t){\rm e}^{i\Delta t/\hbar},\nonumber \\
\langle a_{Xn-1}^{\dagger}(t ) a(t) \rangle&=&\sqrt{n} \rho_{Xn,Xn}(t),\nonumber \\
\langle \varsigma_{n}(t ) a(t) \rangle&=&\sqrt{n} \rho_{Xn,Gn+1}(t){\rm e}^{-i\Delta t/\hbar},
\label{eq:initial1}
\end{eqnarray}
and
\begin{eqnarray}
\langle a_{Gn}^{\dagger}(t ) \sigma(t) \rangle&=& \rho_{Xn,Gn+1}(t){\rm e}^{-i\Delta t/\hbar},\nonumber \\
\langle \sigma_{n}^{+}(t ) \sigma(t) \rangle&=&\rho_{Xn,Xn}(t),\nonumber \\
\langle a_{Xn-1}^{\dagger}(t ) \sigma(t) \rangle&=&0,\nonumber \\
\langle \varsigma_{n}(t ) \sigma(t) \rangle&=&0,
\label{eq:initial2}
\end{eqnarray}
for the calculations of $G_{C}^{(1)}(t,\tau )$ and $G_{X}^{(1)}(t,\tau )$, respectively. 
Here, $\tau$ is calculated up to the stationary limit so that we can compute the two-time correlation functions.


\section*{ACKNOWLEDGMENTS}
We thank Makoto Yamaguchi, Yasutomo Ota, Yukihiro Ota, Kai Yan and Mike Fraser for discussions. 
This work is supported by the Grant-in-Aid for Japan Society for the Promotion of Science (JSPS) Fellows and the Special Coordination Funds for Promoting Science and Technology, the FIRST program for JSPS, Navy/SPAWAR Grant N66001-09-1-2024, Project for Developing Innovation Systems of MEXT, NICT, and Transdisciplinary Research Integration Center. 
This work is also partially supported by the ARO, JSPS-RFBR contract No. 12-02-92100, Grant-in-Aid for Scientific Research (S), and MEXT Kakenhi on Quantum Cybernetics.

\section*{Author contributions}
NI performed calculations. NI and TB wrote the paper. FN and YY supervised the project. All authors reviewed the manuscript.

\section*{Additional information}
Competing financial interests: The authors declare no competing financial interests. 

\end{document}